\documentclass{article}

    \PassOptionsToPackage{numbers, sort&compress}{natbib}

 \usepackage[dblblindworkshop, final]{neurips_2026}
\workshoptitle{NeurIPS 2026 Workshop on Child Safety in AI}

\usepackage[utf8]{inputenc} %
\usepackage[T1]{fontenc}    %
\usepackage[hidelinks]{hyperref}       %
\usepackage{url}            %
\usepackage{xurl}
\usepackage{booktabs}       %
\usepackage{amsfonts}       %
\usepackage{nicefrac}       %
\usepackage{microtype}      %
\usepackage{xcolor}         %
\usepackage{soul}

\usepackage{tikz}
\usetikzlibrary{arrows.meta, positioning}

\title{Open Problem: Protective-Design Pre-Emption in Child-Facing Generative AI}

\author{
Verena Jasmin Hallitschke\thanks{Equal contribution. Correspondence to: \href{mailto:verena-jasmin.hallitschke@uni-tuebingen.de}{\texttt{verena-jasmin.hallitschke@uni-tuebingen.de}}} \\
Hertie Institute for AI in Brain Health \\
T\"ubingen AI Center\\
University of T\"ubingen\\
KI macht Schule\\
\And 
Mahelie Dissanayake\footnotemark[1]\\
T\"ubingen AI Center\\
University of T\"ubingen\\
KI macht Schule\\
\And 
Elisa Nguyen\\
T\"ubingen AI Center\\
University of T\"ubingen\\
KI macht Schule\\
} %

\begin{document}

\maketitle

\begin{abstract}
  Generative AI is adopted rapidly by children, yet unrestricted general-audience models dominate their usage. This paper presents the open problem of \emph{protective-design pre-emption}: Delayed market introduction, low access thresholds, utility gaps, habituation, and peer adoption foster unrestricted AI model lock-in, leading children to bypass child-centric platforms. With first signs of lower adoption of child-centric platforms, we argue that this shift may reinforce existing AI risks such as exposing children to deepened parasocial dependencies, skill erosion, and misinformation. To address this problem, we propose interdisciplinary research directions to quantify pre-emption, understand its socio-technical drivers, and develop interventions ensuring child-centric AI is actively adopted, not just available. 
\end{abstract}

\section{Introduction}
\label{sec:intro}

Multiple independent surveys across countries agree that children indisputably use AI tools for schoolwork, information retrieval, and leisure~\cite{staksrud2026eukidsonline,oneil2026census,unicef2026snapshot,ofcom2026experiences,feierabend2025jim}.
A study of over 25,000 children aged 9 to 16 in the European Union finds that 72\% used generative artificial intelligence (GenAI) in the last month~\cite{staksrud2026eukidsonline}. We distinguish between two categories of AI tools children use. General-audience platforms (e.g., ChatGPT~\cite{openai2022chatgpt}, Gemini~\cite{team2023gemini}) are designed for a broad, mostly adult audience, with safety measures for minors retrofitted after launch, if present at all. Child-centric tools (e.g., Khanmigo~\cite{khan2023khanmigo}) are designed from the outset with age-appropriate content boundaries and child-specific safety guardrails.

There is a clear lag between the introduction of general-audience AI models with new capabilities and the introduction of safe and child-centric variants like Khanmigo~\cite{khan2023khanmigo}. In fact, OpenAI introduced parental controls in ChatGPT almost 3 years after its launch in 2022~\cite{openai2025parentalcontrols} and only started predicting the user age at the beginning of 2026~\cite{openai2026ageprediction}. 
This lag may lead to children using and getting used to popular but unsafe general-audience AI platforms, lowering the adoption of child-centric ones.
We call this \textbf{protective-design pre-emption}: child-centric models are introduced later
such that children come to prefer the unrestricted frontier models they are already using, strengthening a status quo bias.
ChatGPT, as the most popular platform for children by far~\cite{oneil2026census,feierabend2025jim}, highlights the effect. 

We argue that protective-design pre-emption reinforces the risks associated with GenAI usage that child-centric AI is designed to mitigate.
Already, surveys show first signs of reliance on chatbots, with the Common Sense Media census reporting that 42\% of children who use AI daily expect difficulty with not using it for a month~\cite{oneil2026census}.
Children place trust in AI summaries, with almost half of those who use them believing that they are always correct~\cite{ofcom2026children}, highlighting the dangers of lacking AI literacy.
Apart from schoolwork, many children have reported using AI for advice, to practise social skills, and to talk about feelings~\cite{oneil2026census}, opening up potential risks for children's social development.

In this work, we first characterise the open problem of \textbf{protective-design pre-emption} by examining existing statistics on child-centric AI adoption and presenting five contributing factors~(\autoref{sec:problem_desc}).
We further discuss how pre-emption may reinforce existing risks~(\autoref{sec:risk_reinforcement}).
Lastly, we invite the broader community to work on protective-design pre-emption and its downstream effects by proposing concrete research directions~(\autoref{sec:conclusion}).

\section{Characterising Protective-Design Pre-Emption}
\label{sec:problem_desc}

\begin{figure}[t]
    \centering
    \includegraphics[width=0.9\linewidth]{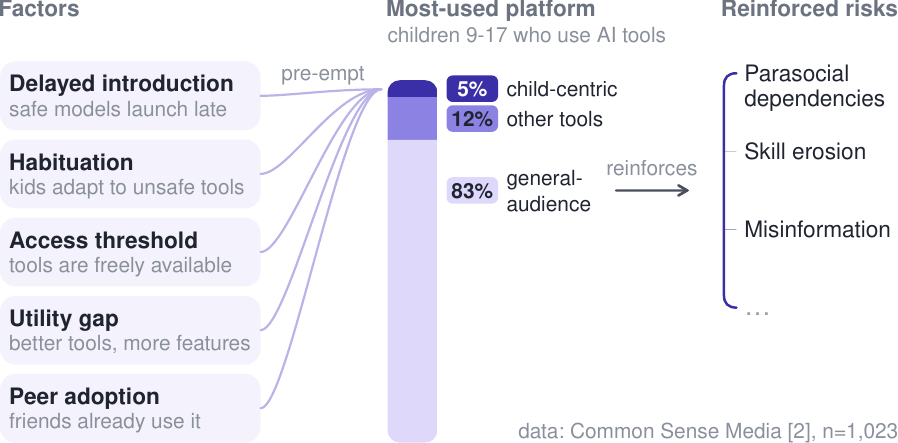}
    \caption{Overview of \textbf{protective-design pre-emption}. We hypothesise five factors contribute to displacing child-centric AI in favour of general-audience tools, reinforcing associated safety risks. The bar shows children's most used AI platforms, consolidated from~\citet[p.53]{oneil2026census} (see~\autoref{sec:consolidation}). The list of risks is not exhaustive, but highlights those illustrated in this work.}
    \label{fig:pdd}
    \vspace{-3mm}
\end{figure}

Protective-design pre-emption is already observable: 
Even where children are provided with a child-centric AI platform, more than half report using other platforms most often, and only 4\% name an educational or school AI platform as their most used one~\cite{oneil2026census}.
Similarly, among children who use AI, 75\% have used AI summaries in search engines, while 38\% report using AI-powered learning tools such as Khanmigo~\cite{oneil2026census}.
This suggests that providing child-centric AI platforms by itself is not sufficient to encourage children to use them, in line with previous works~\cite{robinson2026access,oreopoulos2026one}. %

We argue that five factors contribute to \textbf{protective-design pre-emption} of GenAI platforms: delayed market introduction, habituation, access thresholds, utility gaps, and peer adoption (\autoref{fig:pdd}).

\vspace{-2mm}
\paragraph{Delayed market introduction.}
While general-audience platforms like ChatGPT, introduced in late 2022~\cite{openai2022chatgpt}, are popular with children~\cite{staksrud2026eukidsonline,oneil2026census}, child-centric alternatives such as Khanmigo~\cite{khan2023khanmigo}, ITI~\cite{aileap2026launch}, and AIS.chat~\cite{bremen2025telli} were introduced months or even years afterwards. 
Khanmigo, an AI learning assistant for students and teachers, launched alongside GPT-4 in March 2023, but was initially restricted to users in the United States aged 18 or older~\cite{khan2023khanmigo}.
In Estonia, the AI Leap programme launched in autumn 2025~\cite{aileap2026review}, almost three years later, and students only received access to its learning app ITI in the following January~\cite{aileap2026launch}.
Development of AIS.chat, the federal chat interface for German schools, began in November 2024, and the first federal state introduced it across schools in June 2025~\cite{bremen2025telli}.
Such delayed provision may foster familiarity and pre-emptive lock-in with general-audience platforms and contributes to protective-design pre-emption.

\vspace{-2mm}
\paragraph{Habituation and switching inertia.} 
A study of high school students found that perceived usefulness, social influence, enjoyment, and schoolwork-related use are associated with AI use becoming a habit~\cite{caffaratti2025ai}.
Their general attitude towards AI predicts whether they intend to use it, but not whether that use becomes a habit.
A child-centric alternative might therefore be viewed favourably without changing what children actually use.
Switching costs reduce the intention to switch between GenAI platforms, raise the threshold for dissatisfaction with the current platform, and undermine the experience of alternative platforms~\cite{zhou2025examining}.
We hypothesise that the same applies to children, where established habits could reinforce protective-design pre-emption.

\vspace{-2mm}
\paragraph{Access threshold.} At the time of writing, many general-audience GenAI platforms have low access thresholds. The base versions are free, browser-based, or even built into existing apps, like Meta AI in WhatsApp~\cite{meta2023metaai}, MyAI in Snapchat~\cite{snap2023myai}, or AI summaries in Google~\cite{reid2024aioverviews}.
In contrast, child-centric platforms must be sought out and require adults' permission.
For example, Khanmigo requires a paid subscription for children~\cite{microsoft2024khanacademy}, and access to AIS.chat is granted by teachers who can control students' token use and access duration~\cite{bremen2025telli}. %
Further, these platforms often exclusively cover schoolwork, whereas children also use GenAI for entertainment~\cite{oneil2026census,feierabend2025jim,ofcom2026children,staksrud2026eukidsonline}.
Consequently, general-audience platforms compete with single-purpose, child-centric educational platforms for the same task.

\vspace{-2mm}
\paragraph{Utility gap.}
After the first year, the AI Leap foundation found that students kept using other platforms more frequently than ITI~\cite{aileap2026review}.
Students commonly complained about the number of questions the app asked and switched to other platforms when they were dissatisfied with the responses.
A similar trade-off appears in search, where child-safe filters can impact the relevance of the results~\cite{frobe2025child}, and we hypothesise that safety measures in GenAI could degrade performance in the same way.
We further project that child-centric platforms may adopt better models late enough to fall behind in performance, pushing children towards the platforms that offer the latest models.
The gap extends beyond task performance: adolescents rated a chatbot that disclosed
having no feelings as less likeable and emotionally close than a relational one, and
those preferring the relational style reported poorer relationships and higher
distress~\cite{kim2025iyourelationalconversational}.

\vspace{-2mm}
\paragraph{Peer adoption.}
Social influence is a significant predictor of habitual teenage AI usage~\cite{caffaratti2025ai}. If a child's peers use AI, the child is more likely to follow suit. The parents' stance on AI usage, interestingly, is secondary. A recent study~\cite{bursztyn2026social} makes this clear: parents who do not favour GenAI usage in schools will still pay for AI tools once usage is established within the child's peer group to ensure their child's access to the same technology.
As children learn about GenAI platforms from their parents and peers, among others~\cite{yu2025exploring}, we expect peer adoption to drive AI platform choices as well. 

\section{Reinforcement of Risks as a Consequence of Protective-Design Pre-Emption}
\label{sec:risk_reinforcement}

The pre-emption of child-centric and protectively designed GenAI platforms by general-audience AI platforms threatens to undercut protective mitigations and consequently reinforce existing risks to children. Below, we examine three risks that pre-emption may reinforce as illustrative cases.

\vspace{-2mm}
\paragraph{Parasocial dependencies.}
Children easily form parasocial bonds with non-human agents, as evidenced by normative early-childhood unilateral friendships with imaginary companions~\cite{blake2025parasocial}. Lacking a rigid real/pretend separation, children struggle to assess an AI's aliveness. Conversational AI's reciprocal, personalised responses sustain the illusion of mutual awareness, compounded by an AI \textit{empathy gap}, where systems optimise for affirming dialogue over genuine risk assessment~\cite{kurian2023empathy}.
Children most vulnerable to such bonds may also be those most drawn to unrestricted
systems (\autoref{sec:problem_desc}).

Pre-emption may reinforce these attachments. While child-centric systems feature ``not real'' reminders or escalation prompts~\cite{kurian2025designing}, unrestricted models drive companionship via persistent memory, multimodality and \textit{best friend} marketing~\cite{blake2025parasocial}. 
Notably, 11\% of AI users aged 8 to 17 years report they have used AI as a companion~\cite{ofcom2026children}, and 25\% of children who have used AI to discuss feelings and personal problems feel that the technology understands them better than most people do~\cite{oneil2026census}.
Furthermore, 49\% of children report using AI for advice, 40\% to practise social skills, and 37\% to confide~\cite{oneil2026census}.
The very responsiveness that makes unrestricted systems attractive may also maximise the likelihood of deep, unmitigated parasocial dependencies~\cite{blake2025parasocial}.
Recently, the CARE framework for chatbot design has been proposed~\cite{namvarpour2026understanding}, but its short- and long-term effects on parasocial bond formation have not been studied.
\vspace{-2mm}
\paragraph{Skill erosion.}

We use skill erosion as an umbrella term for \textit{intellectual deskilling} and \textit{skill skipping}.
Relying on generative AI threatens \textit{intellectual deskilling}, which occurs when students rely on technology for tasks they should complete independently~\cite{ivanov2023dark,machindon2025generative, scheiter2026skills}. Recent surveys highlight the scale of this issue: of 2,290 U.S. mothers, only 6\% view AI as positive for children. Conversely, of 2,818 U.S. mothers, 62\% report that AI makes it harder for their children to learn problem-solving and independent thinking~\cite{smales2026children}. 

Pre-emption could worsen this dynamic. While educational AI uses pedagogical guardrails to prevent outsourcing~\cite{machindon2025generative}, migrating to frontier systems that instantly provide direct, high-accuracy answers fosters \textit{skill skipping}. 
This means the non-acquisition of new skills by using AI as a shortcut to obtain direct answers rather than learning the underlying material~\cite{scheiter2026skills}. 
A randomised controlled trial involving nearly 1,000 high school mathematics students exemplifies this threat~\cite{bastani2025generative}. While using a standard AI interface boosted practice performance by 48\%, students overwhelmingly skipped the work, with 56\% of initial interactions simply asking the AI for answers.
When the inclination to skip the work becomes a sustained habit, the resulting deskilling is profound. A longitudinal 30-month study of over 26,811 Chinese students in Grades 7 to 12 quantifies this dynamic~\cite{stromberg2026generative}. The research revealed that general-audience AI boosted homework scores by 18\%, and reduced completion time by 30\%, dropping the average time spent per assignment from 64 to 45 minutes. However, this ultimately caused a 20\% drop in monthly closed-book exam scores within six months.

In the same mathematics trial, a second group practised with a tutor version of the model that was given teacher-provided solutions and instructed to offer hints instead of complete solutions~\cite{bastani2025generative}.
In a subsequent exam without AI access, students who had practised with the unrestricted model scored 17\% lower than a control group without AI, whereas the scores of the guarded group did not differ significantly from the control.
Guardrails can thus mitigate skill erosion, but only for children who actually use the platforms that implement them.

\vspace{-2mm}
\paragraph{Misinformation.}
Conversational AI poses a risk of misinformation because it relies on statistical pattern matching rather than genuine comprehension. Consequently, these models may generate highly confident, ungrounded falsehoods that reflect societal biases~\cite{kurian2023empathy}. Children in particular struggle to separate falsehoods from facts or to recognise when an AI's confidence is unwarranted, as they tend to treat computers as authoritative sources and conflate an AI's confidence with correctness~\cite{kurian2025designing,tian2026}.
The polished responses of these models lead users to perceive them as objective and authoritative~\cite{park2026generative}, and unrestricted models can tailor convincing false narratives that deceive even adults~\cite{park2026generative}.
Among AI-using teenagers, 40\% trust AI-generated news as much as or more than human-written articles, and 46\% of children aged 8 to 17 who read AI search summaries believe they are always accurate~\cite{ofcom2026children}.

While child-centric systems use classifiers and constrained domains~\cite{kurian2025designing}, pre-emption could push children to unrestricted systems lacking these safeguards~\cite{kurian2023empathy}.
A retrieval-augmented tutor restricted to lecturer-provided material answered incorrectly in only 1.5\% of 400 coded responses~\cite{nemeth2025}, illustrating, albeit for university students, what children forgo under pre-emption.
Such restrictions, however, also reflect the utility gap (\autoref{sec:problem_desc}): 16.5\% of the tutor's answers fell outside the provided material~\cite{nemeth2025}, and children aged 10 to 14 perceived reliability and creativity in chatbots as incompatible~\cite{tian2026}.

\section{Conclusion and Research Directions}
\label{sec:conclusion}

In this paper, we shine light on the open problem of \textbf{protective-design pre-emption} as a consequence of customer lock-in to unrestricted, general-audience AI platforms.
We argue that child-centric AI platforms, which already reach few children~\cite{oneil2026census}, are at risk of being displaced further by general-audience ones.
We hypothesise that pre-emption may reinforce already existing risks in child safety regarding GenAI. In fact, we identify indications from existing research that such undesirable reinforcement effects are already apparent, such as parasocial dependencies~\cite{blake2025parasocial}, skill erosion~\cite{stromberg2026generative}, or the spread of misinformation~\cite{ofcom2026children}, among others.

While the child safety risks as well as mitigations like guardrails~\cite{kurian2025designing, yu2025safeguarding, clark2025building} and AI literacy interventions~\cite{su2024artificial,long2020ai,dangol2025children} receive attention in research, we argue that protective-design pre-emption itself receives rather little attention. Yet this effect deserves systematic study in its own right, which opens up several research directions such as: 
\begin{itemize}
    \item \textbf{Quantifying pre-emption and downstream effects:} How often do children revert to general AI chatbots, even when child-centric alternatives exist? To what extent are existing safety risks reinforced? What are the long-term effects of this pre-emption and the implications for child development and society? 
    \item \textbf{Understanding drivers of pre-emption:} What factors contribute to the pre-emption of child-centric AI by general AI? How do child safety features trade off model capabilities and perceptions? How does the usage context, i.e., school or leisure, encourage pre-emption? 
    \item \textbf{Mitigating pre-emption:} How can child-centric, displacement-robust GenAI be designed? What incentive structures would make reducing pre-emption commercially rational for GenAI providers? What strategies can reduce the utility gap and adoption lag for child-centric AI tools, and how can their distinct value proposition relative to general-audience alternatives be communicated more effectively?
    \item \textbf{Regulatory necessity and sufficiency:} How sufficient are measures like age restriction and verification for mitigating underage access to AI platforms? Are there aspects of pre-emption like peer norms or capability gaps that regulation is suited or unsuited to address? 
\end{itemize}

Studying these questions requires collaborative effort across various disciplines such as machine learning, technology policy, law and governance, human-computer interaction, and developmental psychology. We invite the community to study \textbf{protective-design pre-emption} in GenAI to gain insights relevant to how to design AI systems that serve children's and society's best interests. 

\vspace{-2mm}
\paragraph{Limitations.} Our arguments draw on existing statistics and prior literature rather than an original empirical study, and protective-design pre-emption itself remains to be systematically measured. Moreover, the research directions we propose are not exhaustive, but are intended as seed questions for the community to build on.

\begin{ack}
The authors thank the International Max Planck Research School for Intelligent Systems (IMPRS-IS) for supporting EN and VH. This work was supported by the T\"ubingen AI Center. 
We thank the Hertie Foundation, the Carl Zeiss Foundation (CZ Nexus: Certification and Foundations of Safe Machine Learning Systems in Healthcare).
\end{ack}

\bibliographystyle{unsrtnat}
\bibliography{bibliography}

\appendix
\section{Declaration of LLM Usage}
We acknowledge the use of LLM assistants in this work: LLMs were used as writing tools to assist in refining the paper. However, the final version was carefully reviewed and finalised by the authors. No LLMs were used in ideation.

\section{Consolidation of Reported AI Tool Usage} \label{sec:consolidation}

We consolidated the responses from~\citet[p.53]{oneil2026census} to create Figure~\ref{fig:pdd}.
Summing the reported categories assumes that each child was counted under a single response, which the report does not state explicitly.
The group \emph{general-audience} consists of ``ChatGPT'', ``Google'', ``Social media AI'', ``Microsoft Copilot'', ``Virtual assistants'', ``Claude'', and ``AI chatbots'', totalling 71\%.
The group \emph{other AI tools} contains ``Photo, video, or visual design AI tools'', ``Grammarly'', ``Perplexity'', ``AI search'', and ``Other specific AI tools or types'', totalling 10\%.
Finally, the group \emph{child-centric} corresponds to ``Educational AI tools'' with 4\% in the report.
The remaining categories cover children who named no specific tool, who
reported not using AI, or who were not asked.
The number of participants in the report $n=1{,}204$ was scaled by 0.85 to estimate the total number of children that use AI tools (1,023).  
We calculated the share of each AI group by rescaling the aggregated categories to a common base of 100\% and rounded using the largest-remainder
(Hamilton) method so that they sum to 100\%.

\newpage

\end{document}